\documentclass[aps, prl, reprint,twocolumn, preprintnumbers, nofootinbib, showpacs, superscriptaddress]{revtex4-1}
\usepackage{graphicx, amsmath, amssymb, amsfonts, pifont, bm, cancel}
\usepackage[usenames]{color}
\usepackage{subfigure}
\usepackage[normalem]{ulem} 

\definecolor{MyDarkBlue}{rgb}{0,0,1}

 \newcommand{\beq}{\begin{equation}}
 \newcommand{\eeq}{\end{equation}}
 \newcommand{\bel}{\begin{align*}}
 \newcommand{\tamam}{\end{align*}}

 \newcommand{\ket}[1]{|#1\rangle}

 \newcommand{\beqa}{\begin{eqnarray}}             
 \newcommand{\eeqa}{\end{eqnarray}}               
 \newcommand{\bra}[1]{\langle#1\vert}                 

\begin{document}

\title{Wigner distribution of twisted photons}

\author{Mohammad~Mirhosseini}
\email{mirhosse@optics.rochester.edu}
\affiliation{The Institute of Optics, University of Rochester, Rochester, New York 14627, USA}
\author{Omar~S.~Maga\~na-Loaiza}
\affiliation{The Institute of Optics, University of Rochester, Rochester, New York 14627, USA}
\author{Changchen Chen}
\affiliation{The Institute of Optics, University of Rochester, Rochester, New York 14627, USA}
\author{Seyed Mohammad Hashemi Rafsanjani}
\affiliation{The Institute of Optics, University of Rochester, Rochester, New York 14627, USA}
\author{Robert~W.~Boyd}
\affiliation{The Institute of Optics, University of Rochester, Rochester, New York 14627, USA}
\affiliation{Department of Physics, University of Ottawa, Ottawa ON K1N 6N5, Canada}

\date{\today}

\begin{abstract}
{We present the first experimental characterization of the azimuthal Wigner distribution of a photon. Our protocol fully characterizes the transverse structure of a photon in conjugate bases of orbital angular momentum (OAM) and azimuthal angle (ANG). We provide a test of our protocol by characterizing pure superpositions and incoherent mixtures of OAM modes in a seven-dimensional space. The time required for performing measurements in our scheme scales only linearly with the dimension size of the state under investigation. This time scaling makes our technique suitable for quantum information applications involving a large number of OAM states.}
\end{abstract}
\maketitle
Ever since its introduction in 1932 \cite{Wigner:1932cz}, the Wigner distribution has been widely applied in different fields of study ranging from statistical mechanics, and optics \cite{Waller:2012cw} in physics to more applied fields such as electrical engineering and even seismology \cite{Cohen:1989db}. In physics, the Wigner distribution has been utilized to bring the machinery of phase-space statistical mechanics into study of quantum physics \cite{Moyal:1949gj}. Wigner distribution provides a comprehensive characterization of the system, and as a quasiprobability distribution the negativity of the Wigner distribution signals a wave-like behavior \cite{Smithey:1993er,Lvovsky:2001de}.

The orbital angular momentum (OAM) of single photons has lately been identified as a valuable platform for realizing multilevel quantum systems \cite{Allen:1992vk,Mair:2001fd}. The discrete nature of OAM makes it attractive for encoding quantum \cite{Mirhosseini:2015fy} and classical information \cite{Wang2012}. 
The ongoing research suggests that there is no fundamental limit to the maximum value of OAM that a photon can carry. In a recent experiment, quantum entanglement was demonstrated between states differing by 600 in their value of OAM \cite{Fickler:2012hj}. The full characterization of a quantum state in the Hilbert space of OAM poses a serious experimental challenge.

 A large body of previous research has enabled efficient and accurate projective measurements of light's OAM \cite{Mair:2001fd, Gibson2004,Leach2004,Karimi:2009ge,Berkhout2010,Mirhosseini:2013em}. Quantum mechanically, a pure state in the Hilbert space of OAM is described by a discrete state vector. Thus, the probability distribution provided by projective measurements along with the knowledge of relative phase between the different OAM components found by interferometry adequately described a pure state \cite{Malik:2014bf}. Nevertheless, pure states are only a restricted set of physical states, because the vast majority of conceivable states are mixed states \cite{BlumeKohout:2010cm}. The most general description of a quantum state requires knowledge of its density matrix, which can be found through use of standard quantum state tomography \cite{Raymer:1994vt,James:2001bb}. However, quantum state tomography in the OAM basis requires the capability to perform projective measurements on arbitrary superpositions of two or more OAM eigenstates \cite{Bent:2015ej}, a task that remains challenging due to technical limitations such as variations in the efficiency of measuring different OAM modes and the cross-talk between neighboring modes \cite{Qassim:2014fp}.



In this article, we propose and demonstrate a method for obtaining the Wigner distribution for the azimuthal structure of light as an alternative to conventional quantum state tomography. We achieve this task by experimentally finding the projections of the density matrix in the basis of azimuthal angle and subsequently calculating the Wigner function via a linear transformation. This is, to our knowledge, the first experimental characterization of the azimuthal Wigner distribution, a concept that has been a topic of extensive theoretical investigation for the last three decades \cite{Mukunda:1979wx, Dowling:1994tq, Leonhardt:1996ex,Leonhardt:1995ep,Simon:2000uj,Calvo:2005vf,Rigas:2008ez,Rigas:2010ie,Rigas:2011fa}. Our experiment provides valuable insight in understanding the wave behavior of the light field in the conjugate bases of OAM and azimuthal angle, as well as a method for comprehensive characterization of the OAM of single photons that can be used for quantum information applications.

\begin{figure*}
\centerline{\includegraphics[width = 0.8\textwidth]{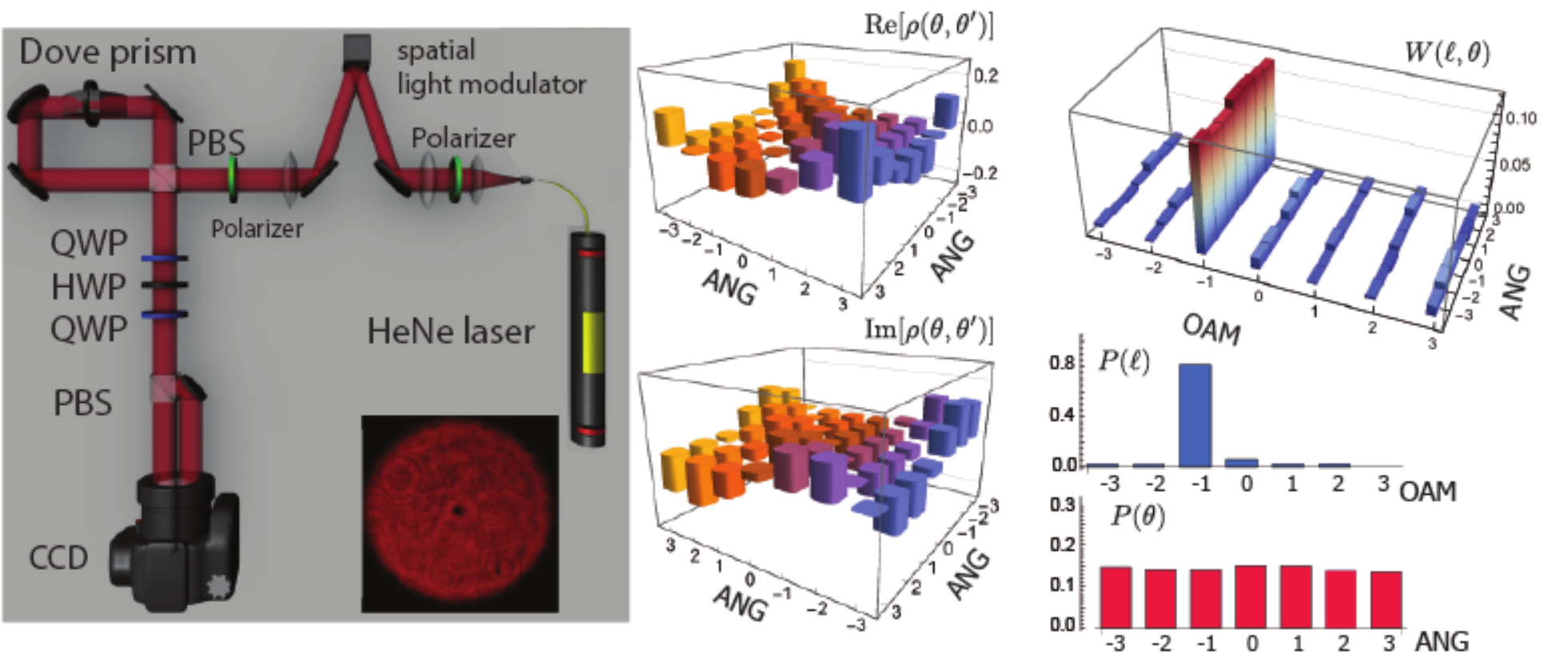}}
\vspace{-3 mm}
\caption{\textbf{Characterization of the transverse structure of classical light.} Left panel: The light beam from a HeNe laser illuminates a phase-only spatial light modulator. The polarization state of the beam is prepared by a polarizer. A Dove prism located inside a Sagnac interferometer causes a rotation in opposite directions of each of the counter-propagating beams. Two quarter-wave plates along with a half-wave are used along with a polarizing beam splitter for characterizing the polarization of the output beam. Middle and right panels: Experimental results for characterization of an OAM mode with $\ell =-1$. The plots in the middle column shows the density matrix in the ANG basis, and the plots in the right column present the azimuthal Wigner distribution along with the corresponding marginal distributions in the ANG and OAM bases.} 
\label{fig:Setup}
\vspace{-5 mm}
\end{figure*}
We begin our analysis by considering a quantum system with an unknown density matrix, $\hat{\rho}$, in the basis of azimuthal angle, $\theta$. Further, we choose to work in a \mbox{finite-dimensional} state space spanned by the orbital-angular-momentum eigenvectors $\ket{\ell}$ with $\{|\ell|\leq N \}$. In this subspace, the (discrete) Wigner distribution function has previously been shown to be \cite{Leonhardt:1996ex, Leonhardt:1995ep}
\begin{align}\label{eq:Wig}
W(\theta, \ell) = \frac{1}{{d}} \sum_{\phi = -N}^{N} \exp{\left(-\frac{4\pi i }{d}\ell \phi \right)}\bra{\theta-\phi}\hat{\rho}\ket{\theta+\phi}.
\end{align}
Here, we have $d = 2N+1$, and $\theta \in \{-N, \ldots, N\}$ denotes the discrete angular coordinate. We have defined an angular (ANG) eigenstate via a discrete Fourier transform of the OAM states
\begin{align}\label{eq:ANG}
\ket{\theta}=\frac{1}{\sqrt{d}}\sum_{\ell= - N}^{\ell= + N} \exp{\left(-\frac{2\pi i}{d} \theta\ell\right)} \ket{\ell}.
\end{align}
Note that the ANG states satisfy the periodicity property,
\vspace{-4 mm}
\begin{align}\label{eq:shift}
\ket{\theta +d } = \ket{\theta},
\end{align}
as expected. The ANG states have previously been introduced in the literature for the purpose of development of angular rotation operators \cite{Barnett:1990do,Leonhardt:1996ex, Leonhardt:1995ep} and also for generalization of the BB-84 QKD protocol to the OAM basis \cite{Mirhosseini:2015fy, Giovannini:2013ju}.

Next, we introduce an ancillary qubit in a different state space, here namely polarization, which is used as a pointer. We assume that the pointer is initially prepared in the state $\ket{+}=(\ket{H}+\ket{V})/\sqrt{2}$, where $\ket{V}$ and $\ket{H}$ stand for vertical and horizontal polarization states. The density matrix associated with the ancila and azimuthal spaces is given by $\hat{\Omega} = \hat{\rho}\otimes\ket{+}\bra{+}$. In the next step, we consider the unitary evolution of the joint system-pointer state characterized by the operator
\vspace{-2 mm}
\begin{align}
\hat{U}(\tau) = \exp{\left(-\frac{2\pi i} {d} \tau \hat{L} \otimes \hat{\sigma}_{z} \right)}.
\end{align}
Here, $\hat{L}$ is the orbital angular momentum operator directed along with the optical axis and $\hat{\sigma}_z = \ket{H}\bra{H}-\ket{V}\bra{V}$, which is one of the Pauli operators for the pointer. Heuristically, the operator $\hat{U}$ describes a polarization-sensitive rotation by the angle $\tau$. After this transformation, the system-pointer state is found as $\hat{\Lambda}(\tau) = \hat{U}^{\dagger}(\tau)\hat{\Omega}\hat{U}(\tau).$ A Hamiltonian of this form has been utilized before for amplifying angular rotations \cite{MaganaLoaiza:2014kf} as well as measuring the moments of OAM \cite{Piccirillo:2015ka}.
 
It is straightforward to verify that the unitary interaction $\hat{U}$ results in an entangled system-pointer state. Post-selection on a specific angular state $\theta$ leads to a reduced density matrix in the Hilbert space of the pointer:
 \begin{align}
&\hat{\sigma} = \frac{\bra{\theta}\hat{\Lambda}\ket{\theta}}{\text{Tr}{\left[\bra{\theta}\hat{\Lambda}\ket{\theta}\right]}}.
\end{align}
\vspace{-2 mm}

We can directly find the elements of density matrix $\hat{\rho}$ by measuring the expectation values of the Pauli operators $\hat{\sigma}_x = \ket{H}\bra{V}+\ket{V}\bra{H}$ and $\hat{\sigma}_y = i\ket{V}\bra{H}-i\ket{H}\bra{V}$ for the pointer. This calculation can be performed by using the shift property of the angular eigenstates, \mbox{$\exp{[-(2\pi i/d) \tau \hat{L}]} \ket{\theta} = \ket{\theta+\tau}$}.
Here, we have $\theta_{\pm}=\theta\pm\tau$. 
Using this notation we find that
\begin{align}\label{eqn:sigma}
\langle\hat{\sigma}_{x}(\theta,\tau)\rangle = \text{Tr}\left[\hat{\sigma}_x \hat{\sigma}\right] = \frac{2}{N(\theta,\tau)}\text{Re}\left[\bra{\theta_{+}}\hat{\rho}\ket{\theta_{-}}\right],\nonumber\\
\langle\hat{\sigma}_{y}(\theta,\tau)\rangle = \text{Tr}\left[\hat{\sigma}_y \hat{\sigma}\right] = \frac{2}{N(\theta,\tau)}\text{Im}\left[\bra{\theta_{+}}\hat{\rho}\ket{\theta_{-}}\right].
\end{align}
Here,  $N(\theta,\tau) = {\text{Tr}{[\bra{\theta}\hat{\Lambda}\ket{\theta}]}}$ is a normalization factor. The pair of equations in Eq.\,(\ref{eqn:sigma}) can be inverted to readily find $\bra{\theta_{+}}\hat{\rho}\ket{\theta_{-}}$. Thus we have found elements of the density matrix in the ANG basis by performing a rotation of value $\tau$, followed by a post-selection $\ket{\theta}$. Note that in this procedure we separately find the real and imaginary parts of the density matrix by measuring the expectation values of the two conjugate variables of the pointer, $\hat{\sigma}_x$ and $\hat{\sigma}_y$. The approach detailed above provides the density matrix in the \mbox{$d$-dimensional} basis of $\ket{\theta}$. Once we find the density matrix in the angular basis, we can use Eq.\,(\ref{eq:Wig}) to find the azimuthal Wigner distribution.

 \begin{figure*}
\centerline{\includegraphics[width = 0.8\textwidth]{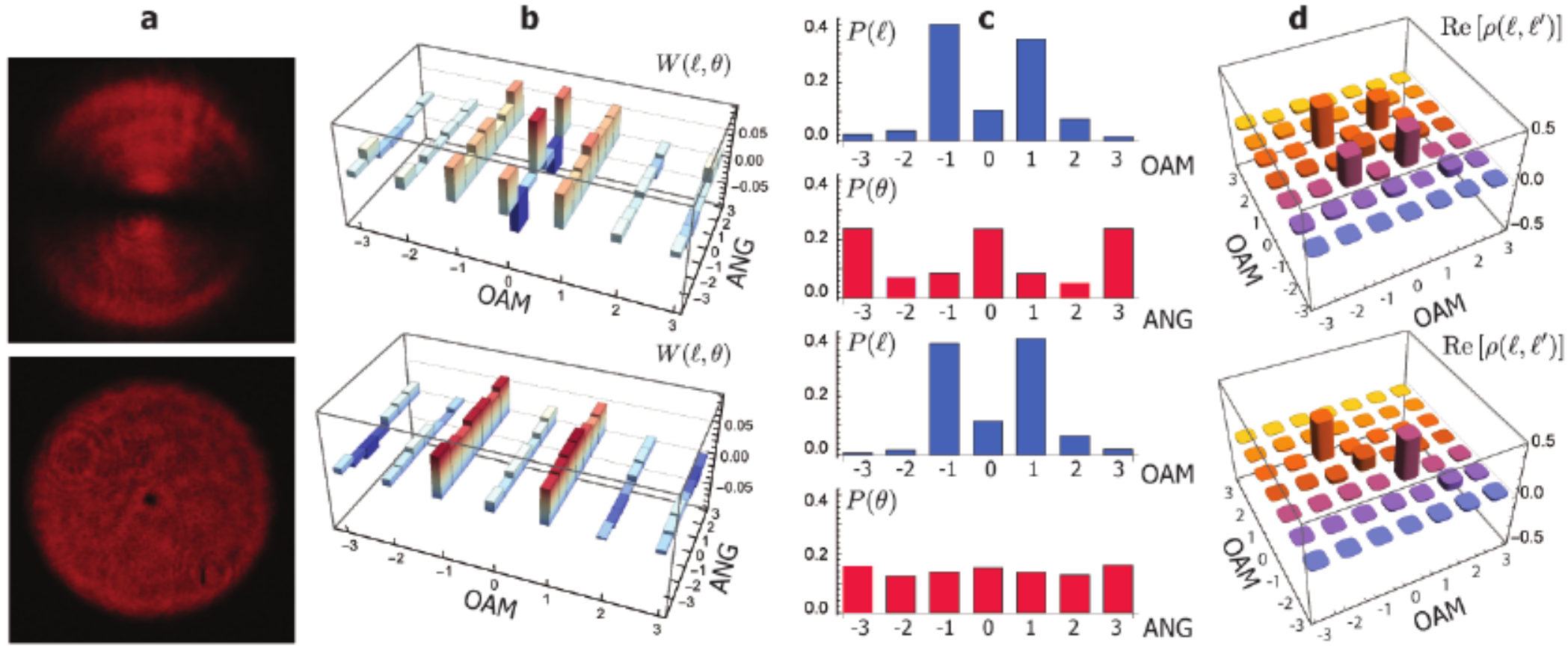}}
\vspace{-3 mm}
\caption{\textbf{Characterization of coherent superposition and a mixture of OAM states.} Left a : The intensity pattern of a pure superposition (top) and (bottom) an incoherent mixture of $\ell = 1$ and $\ell = -1$ OAM modes with equal weights. Panel b : The azimuthal Wigner distribution from the experiment. Panel c : The marginal distributions in the OAM and ANG bases, calculated from the measured Wigner distribution. Panel d : The real and imaginary parts of the OAM density matrix, calculated from the Wigner distribution.} 
\label{fig:Superposition}
\vspace{-5 mm}
\end{figure*}
The left panel of Fig.\,\ref{fig:Setup} illustrates our experimental setup. We use the light beam from a 3 mW He-Ne laser (633 nm), that is coupled to a single-mode fiber (SMF) and then expanded to get a spot size (radius) of 1.8 cm. The laser beam uniformly illuminates the display of the SLM, which has an active area of (9.3 $\times$ 7 mm$^2$ ). The SLM is used to realize computer generated holograms for creating arbitrary spatial modes \cite{Arrizon2007}. We use a Dove prism inside a Sagnac interferometer for realizing the rotational transformation $\hat{U}$. The beam is set to the $45^\circ$ polarization state before the interferometer. We use a quarter- and a half-wave plate along with a polarizing beam splitter (PBS) for realizing the measurement of $\langle\hat{\sigma}_x\rangle$ and $\langle\hat{\sigma}_y\rangle$.

It is possible to experimentally realize projection onto angular states defined in Eq.\,(\ref{eq:ANG}) with a series of custom optical elements \cite{Mirhosseini:2013em, OSullivan:2012gj}. However, post-selection on an angular wedge with sharp boundaries is a much simpler task that provides all necessary information for finding the density matrix in the ANG basis. We achieve this task by recording the intensity of the beam at the two output ports of the PBS with a charge-coupled device (CCD) camera. Once we record the intensity in form of an image, it can be binned to a sequence of numbers that correspond to post-selection on multiple angular states. In the supplementary material we have detailed the process of converting measurement results onto the elements of density matrix in the ANG basis.

To confirm our characterization method, we test it on a series of different states. Figure\,\ref{fig:Setup} also shows experimental results for the characterization of an $\ket{\ell =1}$ OAM mode generated by the SLM. The middle panel shows the real and imaginary parts of the density matrix in the ANG basis, directly found from the experiment.  The right panel demonstrates the Wigner distribution in the conjugate bases of OAM and ANG. The marginals show the probability distributions of the state in the OAM and ANG bases. It is evident that the state primarily constitutes the $\ket{\ell =1}$, and that it includes (approximately) equal components of ANG states.
We calculate the reasonably high fidelity of the characterized state with $\ket{\ell =1}$ as $90\%$, testifying to the high quality of the generation and the characterization procedure. We have used the standard method of maximum-likelihood estimation to find a positive-definite density matrices in the ANG basis from the experimental data \cite{Thew:2002tc}.

\begin{figure*}
\centerline{\includegraphics[width = 0.8\textwidth]{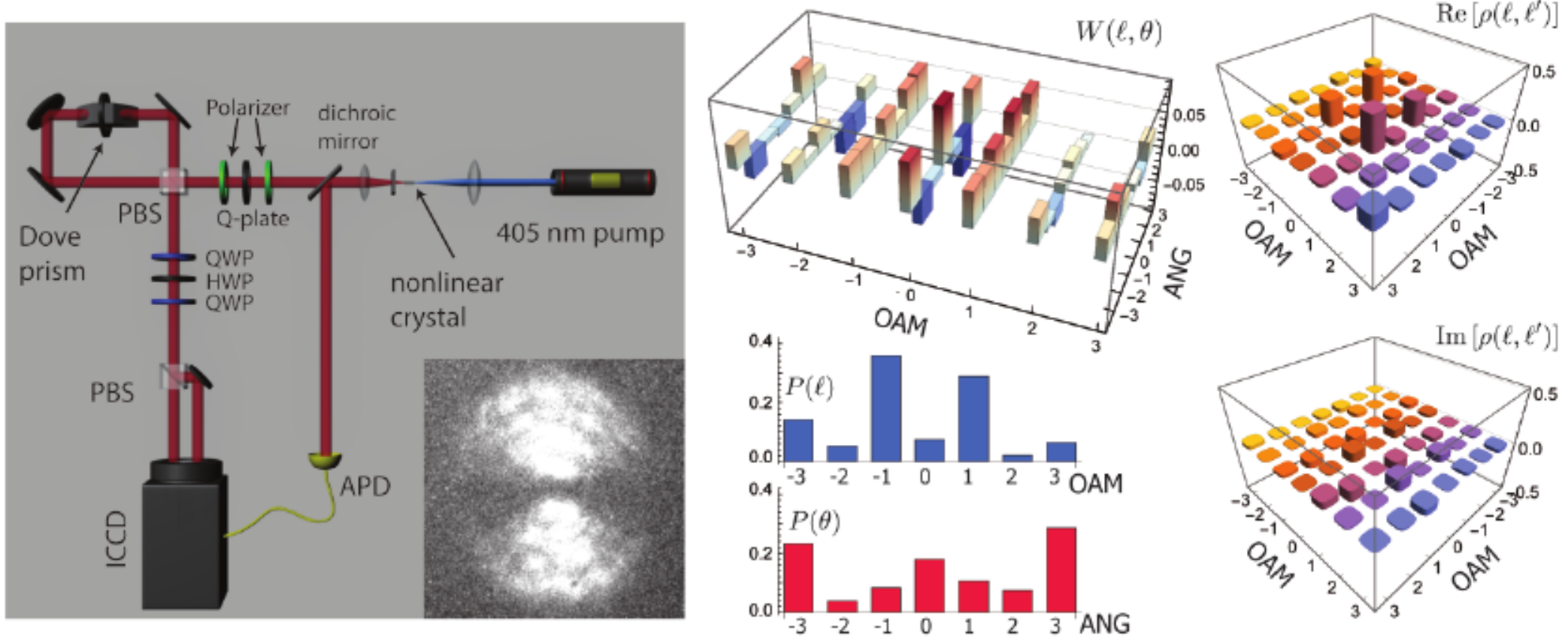}}
\vspace{-3 mm}
\caption{\textbf{Characterization of the transverse structure of single photons.} Left panel: A PPKTP crystal is pumped with a 405 nm continuous wave laser beam. Single photons from non-degenerate parametric down-conversion are separated ny a dichroic mirror. The idler photons (830 nm) are detected by an avalanche photo-diode (APD), which heralds the detection of signal photons (790 nm) with an intensified charge coupled device (ICCD). A $q$-plate (q = $1/2$) is placed between two crossed polarizer to prepare an equal superposition of $\ell = 1$ and $\ell = -1$ OAM modes. Inset: The transverse structure of single photons captured with an accumulation of 5-ns-coincidence events over a 1200 sec exposure time. Right panel: The Wigner distribution, the OAM and ANG marginals, and the real and imaginary parts of the OAM density matrix from experiment.} 
\label{fig:QuantSetup}
\vspace{-5 mm}
\end{figure*}
As another test, we generate and characterize an equal superposition of the OAM states $\ket{\ell = 1}$ and $\ket{\ell = -1}$. A pure superposition state is generated directly through the use of a computer generated hologram. To create a mixed state, we use a computer program to randomly switch the SLM between two holograms designed for generating $\ell =1$ and $\ell = -1$ modes. The mode switching occurs at a rate of $60$ Hz, and we use a long ($10 $ s) exposure time on the CCD to guarantee uniform averaging over the changing beam structure. Figure\,\ref{fig:Superposition} shows the intensity patterns and the measured Wigner distributions for the two states. It is evident that marginal distributions in the OAM bases are nearly identical, demonstrating the two prominent contributions from $\ket{\ell = 1}$ and $\ket{\ell = -1}$ in both cases. However, the Wigner distributions and the marginal distributions in the ANG bases are entirely different. For the pure superposition, we observe an interference pattern in the ANG marginal, and negative values on the $\ket{\ell = 0}$ portion of the Wigner distribution. For the incoherent mixture, we see no interference in the ANG marginals, and and the $\ket{\ell = 0}$ portion of the Wigner distribution remains positive. This is a manifestation of a well known property of the Wigner distribution. Namely, wave interference gives rise to negative values on the Wigner distribution, whereas such a pattern is absent for an incoherent mixture.

We have mapped the Wigner distribution onto the OAM density matrix for the states presented in Fig.\,\ref{fig:Superposition}. The degree of coherence between the OAM components $\ket{\ell = 1}$ and $\ket{\ell = -1}$ can now be quantified by the magnitude of the off-diagonal elements of the density matrix. We calculate the degree of coherence using the relation
\begin{align}
\gamma = \frac{|\rho{(-1,1)}|}{\sqrt{|\rho{(1,1)}||\rho{(-1,-1)}|}}.
\end{align}
We find the degree of coherence for the two states under consideration as $\gamma_{\text{pure}} = 0.80 $ and $\gamma_{\text{mixed}} = 0.06 $. For the pure superposition state, we attribute the slight reduction from unity of the degree of coherence to the imperfections in the generation of the state and the averaging over the non-uniform radial structure of the laser beam. In addition to the results presented above, we have tested our method on a number of different states in the angular and OAM bases (see the supplementary information).

The high photon efficiency of our method makes it suitable for characterization of quantum sources of light, which are often severely limited in the photon flux. We test our method by characterizing the transverse structure of heralded single photons using the setup depicted in Fig.\,\ref{fig:QuantSetup}. We generate pairs of photons by pumping a periodically poled potassium titanyl phosphate crystal (PPKTP) with the beam from a 405 nm laser diode \cite{Steinlechner:2012ex}. The type-0 parametric down-conversion process converts a photon of the pump beam to a pair of signal and an idler photons at the wavelength of 790 nm and 830 nm respectively. We separate the two photons of each pair by using a dichroic mirror. The idler photons are then collected with a lens and detected using an avalanche photo-diode (APD). The signal photons are sent through a $q$-plate that is sandwiched between two crossed polarizers. We use a $q$-plate with a charge of $1/2$ to shape the transverse structure of the photon to a superposition of $\ket{\ell = 1}$ and $\ket{\ell = -1}$ states \cite{Marrucci:2006ga}. The structured photons are sent through the Sagnac interferometer described above. We use an Andor iStar intensified charge coupled device (ICCD) camera for detecting the heralded single photons \cite{Fickler:2013go}. Each detection event is triggered by the electronic signal from the APD in a 5 ns time window. Figure\,\ref{fig:QuantSetup} displays the structure of the shaped signal beam from a 1200 sec exposure. We combine our measurement results for the different rotation angles to find the Wigner distribution and subsequently map it to the OAM density matrix (see the middle and right panels of Fig.\,\ref{fig:QuantSetup}). The Wigner distribution exhibits regions of substantial negative value for $\ell = 0$ portion, which demonstrated quantum interference between $\ell = 1$ and $\ell = -1$ components of the state. In addition, it is evident that the density matrix closely resembles the one from measurement of a $\ket{\ell = 1}$ and $\ket{\ell = -1}$ previously performed with a classical beam of light (for comparison refer to Fig.\,\ref{fig:Superposition}). 

We conclude our remarks by analyzing the scaling of our characterization technique. For the full characterization of the density matrix in a Hilbert space of dimension $d= 2N+1$, one needs to measure $d^2-1$ uknown quantities \cite{James:2001bb}. The quadratic scaling of the number of required measurement has posed a long-standing challenge for measuring states with large dimensions \cite{Agnew:2011jsa, Mirhosseini:2014ch}. Through the use of a CCD/iCCD camera for post-selection, we are able to sequence individual images to find $d$ elements of the density matrix simultaneously. This is a crucial practical advantage since our measurement time scales linearly (as apposed to quadratically) with the dimensionality of the state. Considering the values of exposure times and the resolution of the CCD/iCCD cameras used in this work, we anticipate that the characterization of state with a dimensionality, $d$, as large as 100 is feasible with the demonstrated technique.

In summary, we have demonstrated a technique for the full characterization of the azimuthal structure of a photon. We have achieved this task by finding the azimuthal Wigner distribution via projections in the angular basis. We have used a linear transformation to map the Wigner distribution onto the OAM density matrix. We have tested our technique by applying it to the characterization of both classical laser beams and heralded single photons. Our approach readily scales to very large dimensions, involves no photon loss from post-selection, and is capable of characterizing partially coherent OAM states. To our knowledge, this technique is the only approach that is capable of simultaneously achieving these goals. We anticipate that the presented method for characterization of the azimuthal Wigner distribution will constitute an essential part of quantum information protocols that employ the azimuthal structure of single photons.\\
We acknowledge Ebrahim Karimi for providing the $q$-plate, and for helpful discussions.\\

\bibliographystyle{apsrev4-1}
\bibliography{WOAM}

\newpage
\section{Supplementary material}

\subsection*{Measurement in the wedge basis}
We have previously defined the angular (ANG) states as 
\begin{align}\label{eq:ANG}
\ket{\theta}=\frac{1}{\sqrt{d}}\sum_{\ell= - N}^{\ell= + N} \exp{\left(-\frac{2\pi i}{d} \theta\ell\right)} \ket{\ell}.
\end{align}
To approximate the ANG states, we define the wedge states as
\begin{align}
\ket{\Theta}=  \frac{\sqrt{{2}{\pi}}}{d}\sum_{\ell = -\infty}^{\infty}{\text{sinc}{\left(\frac{\ell \pi}{d} \right)}} {\exp{\left(-\frac{2\pi i}{d} \Theta \ell\right)}} \ket{\ell}.
\end{align}
Here, we have $\Theta = \{-N:1:N\}$, and $\text{sinc}(x) = {\sin(x)}/{x}$. It is easy to check that the wedge states defined above possess the same mean angular positions as the ANG states for $\Theta = \theta$, and two wedge states with different values of $\Theta$ are orthogonal to each other. We use capital Greek letters for denoting the wedge states, and lower case Greek letter for the ANGs to avoid confusion. 

The shift property of the ANG modes plays a crucial property in our analysis. It is straightforward to show that the wedge states $\ket{W_n}$ satisfy this property as well. Namely, we have
\begin{align}
\exp{\left(-\frac{2\pi i}{d} \Omega \hat{L}\right)} \ket{\Theta} = \ket{\Theta+\Omega}.
\end{align}

Using the shift property and following the analysis we previously developed, we find the projections of the density matrix in the wedge basis as
\begin{align}
\bra{\Theta_{+}}\hat{\rho}\ket{\Theta_{-}} =\frac{N(\Theta,\Omega)}{2} \left[ \langle\hat{\sigma}_{x}(\Theta,\Omega)\rangle + i  \langle\hat{\sigma}_{y}(\Theta,\Omega)\rangle \right].
 \end{align}
 Here, we have $\Theta_{+} = \Theta+\Omega$ and $\Theta_{+} = \Theta-\Omega$. 
 
A post-selection onto a wedge state can be performed simply by passing a beam of light through an angular slit. Thus, it is advantageous to use the wedge basis in the experiment as an alternative to the ANG basis. However, we need to find the projections of the density matrix in the ANG basis in order to find the Wigner distribution function. Thus, the question arise whether the projection results in the wedge basis are sufficient to find the elements of density matrix in the ANG basis. We show that this is in fact possible by providing a procedure for achieving this basis conversion.

A straightforward basis conversion would be possible if the ANG states could be written as a superposition of the wedge states. Nevertheless, the wedge states reside in a larger Hilbert space as compared to that of the ANG states. This is evident from the OAM spectrum of the wedge states, which spans $\ell = -\infty$ to $\ell = +\infty$. We define a new set of states as

\begin{align}
\ket{\overline{\Theta}}=  \frac{\sqrt{{2}{\pi}}}{d}\sum_{\ell = -N}^{N}{\text{sinc}{\left(\frac{\ell \pi}{d} \right)}} {\exp{\left(-\frac{2\pi i}{d} {\Theta} \ell\right)}} \ket{\ell},
\end{align}
We call these states as modified wedge (MW) states. It is evident that in the OAM basis, a MW state is identical to its corresponding wedge with for the range of OAM modes $|\ell|\leq N$. Unlike the wedge states, however, the MW states do not have any components outside this range. Because of this property, the MW states reside in the same subspace of the Hilbert space as the ANG modes and the OAM modes under consideration. 
 
 It is straightforward to show that the OAM states in the range $|\ell|\leq N$ can be written as a superposition of the MW modes. To show that this expansion is indeed possible, we evaluate the following expression
  \begin{align}
 &\sum_{\Theta = -N}^{N}  \exp{\left(\frac{2\pi i}{d} \Theta \ell\right)} \ket{\overline{\Theta}} \nonumber\\ &= \frac{\sqrt{{2}{\pi}}}{d}  \sum_{\Theta = -N}^{N}\sum_{\ell' = -N}^{N}{\text{sinc}{\left(\frac{\ell' \pi}{d} \right)}} {\exp{\left[-\frac{2\pi i}{d} {\Theta} (\ell-\ell')\right]}} \ket{\ell'} \nonumber\\
 &= \frac{\sqrt{{2}{\pi}}}{d}  \sum_{\ell' = -N}^{N}{\text{sinc}{\left(\frac{\ell' \pi}{d} \right)}} \ket{\ell'} \sum_{\Theta = -N}^{N}{\exp{\left[-\frac{2\pi i}{d} {\Theta} (\ell-\ell')\right]}}\nonumber\\
&= \sqrt{2\pi}  \sum_{\ell' = -N}^{N}{\text{sinc}{\left(\frac{\ell' \pi}{d} \right)}} \ket{\ell'} \delta_{\ell,\ell'} \nonumber\\
&= \sqrt{2\pi}~{\text{sinc}{\left(\frac{\ell \pi}{d} \right)}} \ket{\ell}.
\end{align}
 
 We now expand an OAM state as 
\begin{align}
\ket{\ell} = \frac{1}{\sqrt{2\pi}} \sum_{\Theta = -N}^{N} \frac{(\frac{\ell \pi}{d})}{\sin{(\frac{\ell \pi}{d})}} \exp{\left(\frac{2\pi i}{d} \Theta \ell\right)} \ket{\overline{\Theta}}, 
\end{align}
for $|\ell|\leq N$. Subsequently, an ANG mode can be expanded in the MW basis
\begin{align}
\ket{\theta} = \sum_{\Theta = -N}^{N} C_{\theta, \Theta} \ket{\overline{\Theta}}.
\end{align}
Here, the expansion coefficients, $a_{\theta, \Theta}$, are found as
\begin{align}
C_{\theta, \Theta} = \frac{1}{\sqrt{2\pi d}} \sum_{\ell = -N}^{N} \frac{(\frac{\ell \pi}{d})}{\sin{(\frac{\ell \pi}{d})}} \exp{\left[\frac{2\pi i}{d} \ell (\Theta- \theta)\right]}.
\end{align}

Having found an expansion for the ANG modes, we can readily expands the elements of the density matrix in the ANG basis in terms of the density matrix elements in the MW basis 
\begin{align}
\bra{\theta_{+}}\hat{\rho}\ket{\theta_{-}} =  \sum_{\Theta = -N}^{N}\sum_{\Theta' = -N}^{N} C_{\theta, \Theta'} C^*_{\theta, \Theta} \bra{\overline{\Theta}}\hat{\rho}\ket{\overline{\Theta'}}.
\end{align}
Here, $\theta_{+} = \theta + \tau$ and $\theta_{-} = \theta - \tau$. Since we have assumed our state of interest, $\rho$, resides in the space spanned by $|\ell|\leq N$, a projection of the density matrix onto a pair of wedge states provides identical results to a projection onto a pair of MWs. Namely, we have
\begin{align}\label{eq:WtoME}
\bra{\Theta}\hat{\rho}\ket{\Theta'} = \bra{\overline{\Theta}}\hat{\rho}\ket{\overline{\Theta'}}, 
\end{align}
 for all values of $\Theta$ and $\Theta'$. Using this result, we now rewrite Eq.\,(\ref{eq:WtoME}) in terms of the wedge basis projections
\begin{align}
\bra{\theta_{+}}\hat{\rho}\ket{\theta_{-}} =  \sum_{\Theta = -N}^{N}\sum_{\Theta' = -N}^{N} C_{\theta_{+}, \Theta'} C^*_{\theta_{-}, \Theta} \bra{{\Theta}}\hat{\rho}.\ket{{\Theta'}}.\end{align}
Thus, we have provided a transformation to find the density matrix element in the ANG basis, $\bra{\theta_{+}}\hat{\rho}\ket{\theta_{-}}$, from our measurement results in the wedge basis $\bra{{\Theta}}\hat{\rho}\ket{{\Theta'}}$. The density matrix elements in the ANG basis can be subsequently used to calculate the Wigner distribution function using the relation
\begin{align}\label{eq:Wig}
W(\theta, \ell) = \frac{1}{{d}} \sum_{\phi = -N}^{N} \exp{\left(-\frac{4\pi i }{d}\ell \phi \right)}\bra{\theta-\phi}\hat{\rho}\ket{\theta+\phi}.
\end{align}

\subsection*{Additional Laboratory Results}

Below, we present experimental results for the characterization of a number of states. Figures 1 through 4 present additional laboratory results.  Fig 1 shows results for characterization of a coherent superposition and a mixture of wedges. Figure 2 shows results for characterization of OAM eigenstates. Figure 3 shows results for characterization of a coherent superposition and a mixture of OAM eigenstates, and Figure 4 shows results for characterization of a coherent superposition of OAM eigenstates imposed on the transverse structure of single photons
.
 \begin{figure*}[h]
\centerline{\includegraphics[width = \textwidth]{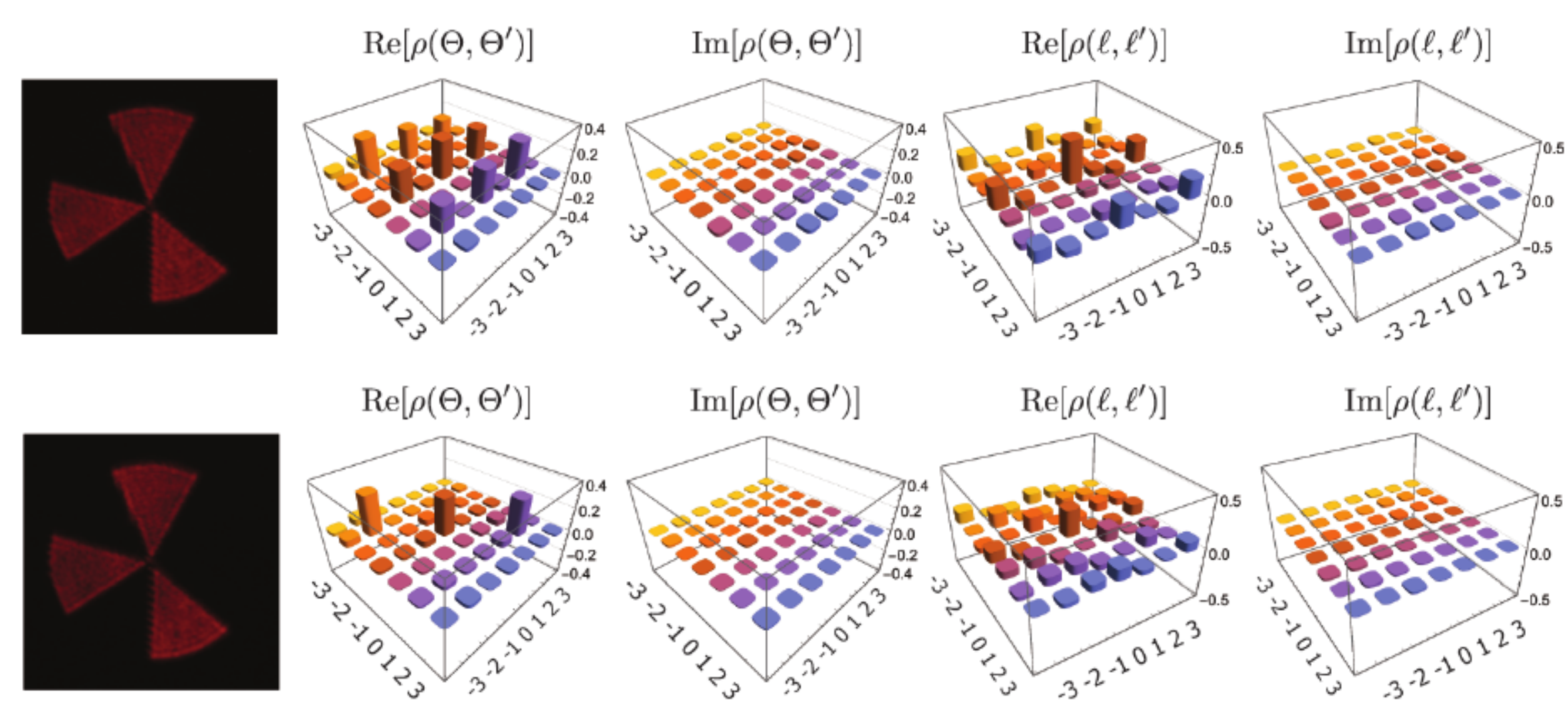}}
\caption{\textbf{Characterization of a coherent superposition and a mixture of wedges.} Top panel: Elements of the density matrix for  a coherent superposition of three wedges ($\ket{\Psi} = \ket{\Theta_1} + \ket{\Theta_3} + \ket{\Theta_5}$). Bottom panel: Elements of the density matrix for an incoherent mixture of the same wedge states.}
\end{figure*}

 \begin{figure*}
\centerline{\includegraphics[width = 1\textwidth]{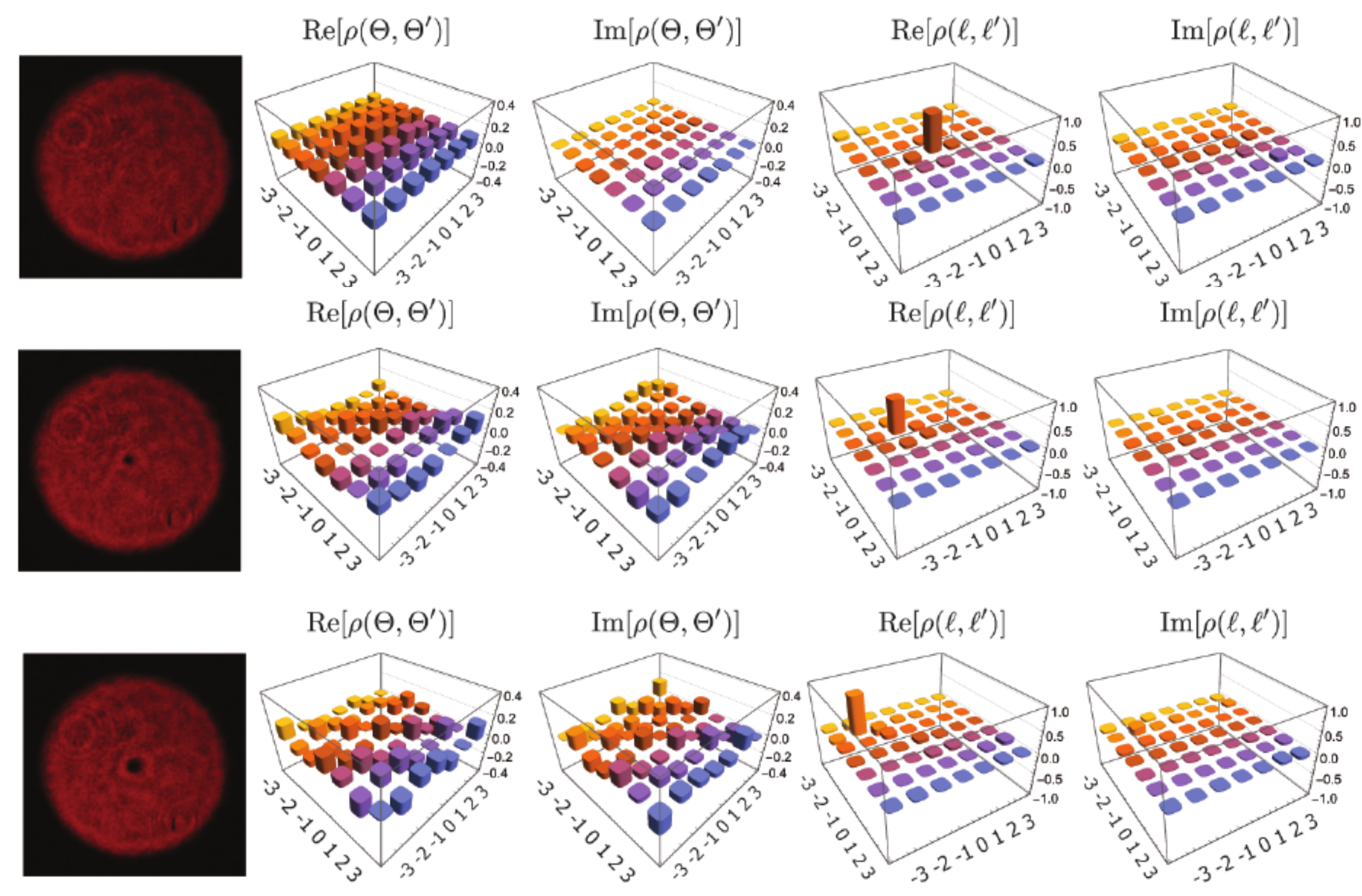}}
\caption{\textbf{Characterization of OAM eigenstates.} Top panel: Elements of the density matrix for $\ell =0$. Middle panel: Elements of the density matrix for $\ell = -1$. Bottom panel: Elements of the density matrix for $\ell = - 2$.}
\end{figure*}

\clearpage

\begin{figure*}
\centerline{\includegraphics[width = 1\textwidth]{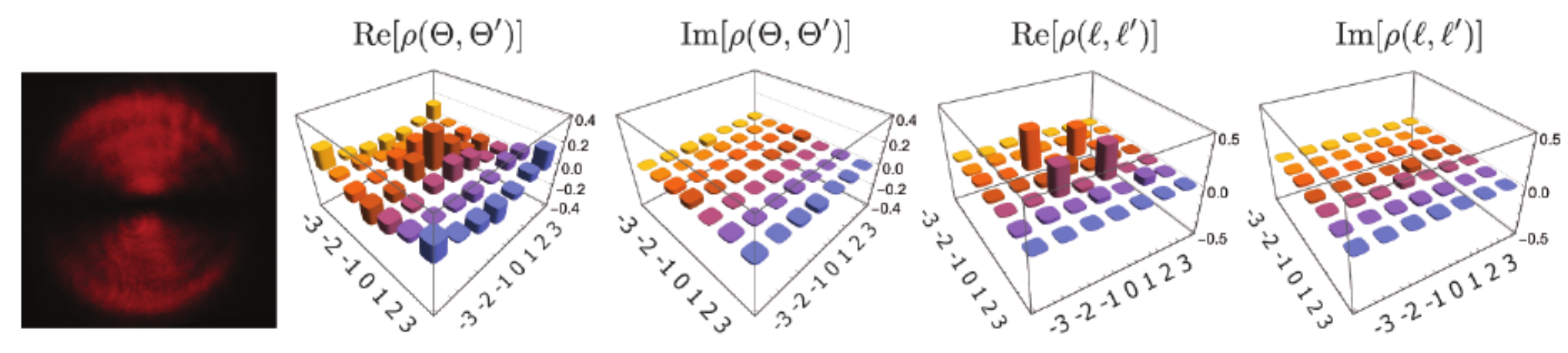}}

\centerline{\includegraphics[width = 1\textwidth]{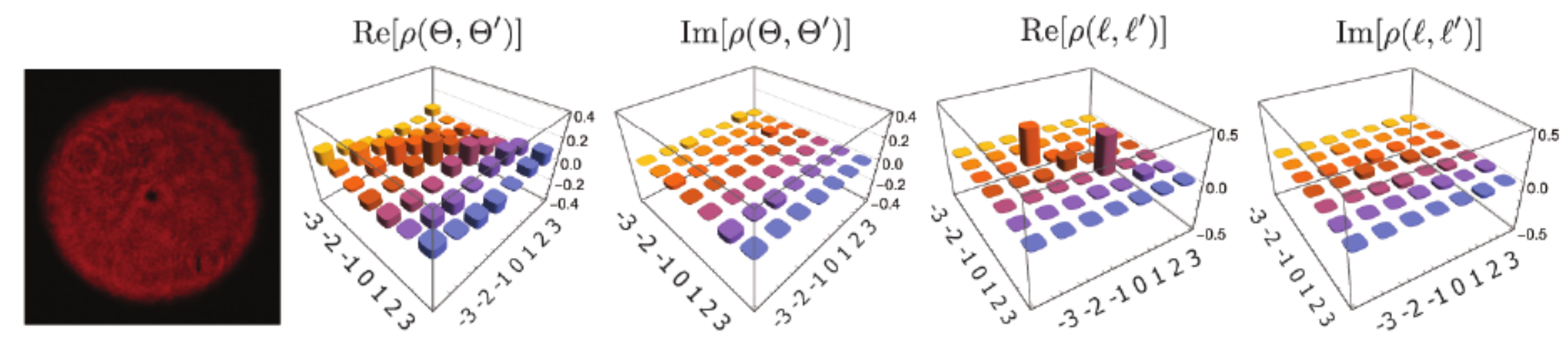}}
\caption{\textbf{Characterization of a coherent superposition and a mixture of OAM eigenstates.} Top panel: Elements of the density matrix for a coherent superposition of two OAM states ($\ket{\Psi} = \ket{\ell} + \ket{-\ell}$, where $\ell = 1$). Bottom panel: Elements of the density matrix for an incoherent mixture of two OAM states ($\rho = \ket{\ell}\bra{\ell} + \ket{-\ell}\bra{-\ell}$, where $\ell = 1$).}

\end{figure*}

%
%

\begin{figure*}
\centerline{\includegraphics[width = 1\textwidth]{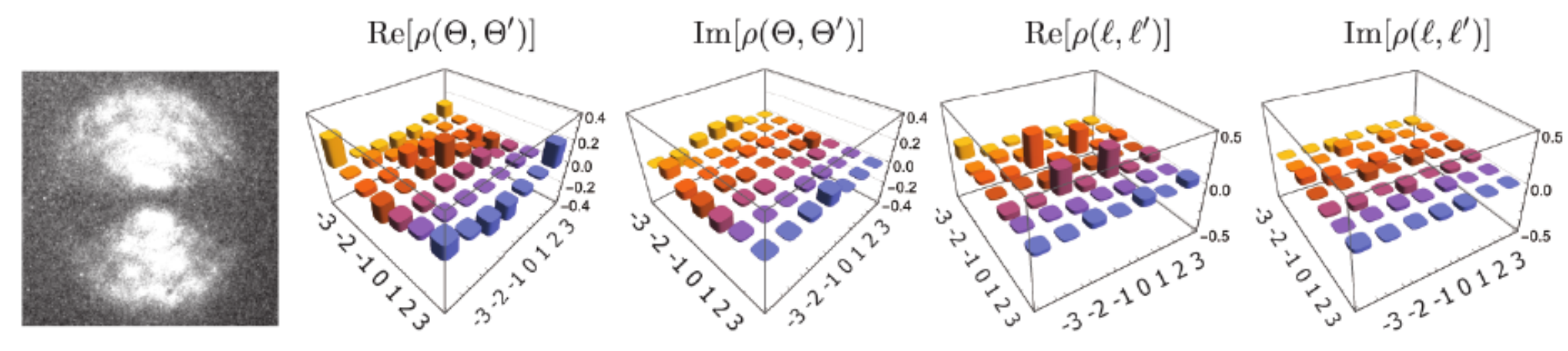}}
\caption{\textbf{Characterization of a coherent superposition of OAM eigenstates imposed on the transverse structure of single photons.}  }
\end{figure*}

\end{document}